\documentclass{elsarticle}

\usepackage{amsmath,amstext,amssymb}
\usepackage{psfrag,subfigure,xcolor,color}
\usepackage{graphicx}
 
\def\caln{{\cal N}}

\def\calc{{\cal C}}
\def\ie{{i.\,e.\ }}
\def\eg{{e.\,g.\ }}
\def\dt{\tilde d}
\def\mut{\tilde\mu}
\def\ft{\tilde f}
\def\vrho{\varrho}
\def\dt{\tilde d}
\def\At{\tilde A}
\def\pt{\tilde p}
\def\w{\mathfrak{w}}

\def\calw{{\cal W}}

\def\re{\text{Re}\:}

\def\del{\partial}

\def\str{\mathrm{Str}}

\def\ii{{\mathrm i}}

\newcommand{\dd}{\mathrm{d}}

\begin{document}


\title{Superconductivity from gauge/gravity duality with flavor}
\date{\today}

\cortext[cor1]{Corresponding Author}

\author[MPI]{Martin Ammon}
\ead{ammon@mppmu.mpg.de}
\author[MPI]{Johanna Erdmenger}    
\ead{jke@mppmu.mpg.de}
\author[MPI,UAM]{Matthias Kaminski} 
\ead{kaminski@mppmu.mpg.de}
\author[MPI]{Patrick Kerner\corref{cor1}}
\ead{pkerner@mppmu.mpg.de}

\address[MPI]{Max-Planck-Institut f\"ur Physik (Werner-Heisenberg-Institut)\\
  F\"ohringer Ring 6, 80805 M\"unchen, Germany}
\address[UAM]{
 Instituto de Fisica Te\'orica UAM/CSIC
 Facultad de Ciencias, C-XVI
 Universidad Aut\'onoma de Madrid
 Cantoblanco, Madrid 28049, Spain

}



\begin{abstract}
\noindent 
We consider thermal strongly-coupled $\caln=2$ SYM theory with fundamental
matter at finite isospin chemical potential.  Using gauge/gravity
duality, \ie a probe of two flavor D$7$-branes embedded in the AdS black hole
background, we find a critical temperature at which the system undergoes a 
second order phase transition. The critical exponent of this transition 
is one half and coincides with the result from mean field theory. In the 
thermodynamically favored phase, a flavor current acquires a vev and breaks an
Abelian symmetry spontaneously. This new phase shows signatures known from superconductivity, such as an
infinite dc conductivity and a gap in the frequency-dependent
conductivity. The gravity setup allows for an explicit identification of the
degrees of freedom in the dual field theory, as well as for a dual string
picture of the condensation process.
\end{abstract}


\maketitle

Gauge/gravity duality is a powerful tool to compute 
observables of field theories at strong coupling. The original 
duality for superconformal field theories has
been extended to field theories at 
finite temperature~\cite{Witten:1998zw}, using an AdS black hole as
gravity dual. Moreover, the original correspondence has been extended
to field theories with flavor degrees of freedom in the fundamental
representation of the gauge group and to finite densities.
In~\cite{Erdmenger:2008yj}, a finite $SU(2)$ 
isospin density $\tilde d$ 
was considered. At high isospin densities $\tilde d>\tilde d_c$, 
this configuration becomes unstable against 
flavor current fluctuations, corresponding 
to vector mesons~\cite{Erdmenger:2007ja}. 
\begin{figure}
  \centering
  \includegraphics[width=0.6\linewidth]{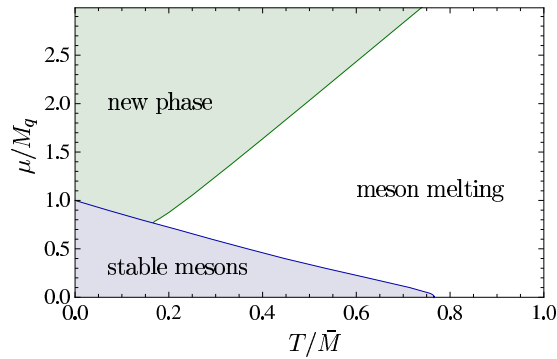}
  \caption{Phase diagram for fundamental matter in thermal strongly-coupled
    $\caln=2$ SYM theory~\cite{Erdmenger:2008yj}, with $\mu$ the
isospin chemical potential, $M_q$ the bare quark mass, $\bar M = 2 M_q \lambda^{-1/2}$,  $\lambda$ the 't Hooft coupling and $T$ the temperature: 
In the blue shaded region, mesons are stable. 
In the white and green regions, the mesons melt. Here the new phase is
stabilized while it was unstable in~\cite{Erdmenger:2008yj}. In this phase we find some features known from superfluids.} 
  \label{fig:phasediagram}
\end{figure}

The central result of the present letter is that there is a new stable
phase characterized by a $\vrho$-meson superfluid.  We show explicitly that 
the new phase shown in fig.~1 is thermodynamically preferred. Moreover, we
check that the phase is stable (at least against the fluctuations which
caused the instability beforehand).

Moreover we find that this phase has properties of a superconductor, such as a second order
phase transition with a critical
exponent of 1/2. The flavor current, which is analogous to the
electromagnetic current, displays infinite dc conductivity and a gap in the
frequency-dependent conductivity. Furthermore, we give a dual
string-theoretical picture of the Cooper pairs and find a dynamical generation
of meson masses similar to the Higgs mechanism. The stabilization mechanism is
motivated by the results of~\cite{Buchel:2006aa,Aharony:2007uu}. In the
context of a phenomenological AdS/QCD model,
a similar stabilization mechanism was first used
in~\cite{Gubser:2008wv}
to obtain a p-wave superconductor.

Our results provide an important link between
gauge/gravity with flavor and the recent gauge/gravity realizations of
superconducting states  \eg in
\cite{Gubser:2008wv,Hartnoll:2008vx,Hartnoll:2008kx}, as well as
a realization of a superconducting current from a top-down
(super-) gravity model, rather than in a phenomenological bottom-up AdS/QCD
model.  Moreover, the degrees of freedom in the dual field theory are
identified explicitly.

\section{FIELD THEORY INTERPRETATION}

In this letter we consider 3+1-dimensional $\mathcal{N}=2$ supersymmetric
Yang-Mills theory, consisting of a $\mathcal{N}=4$ gauge multiplet as well
as two $\mathcal{N}=2$ supersymmetric hypermultiplets, called $(\phi_u,
\psi_u)$ and $(\phi_d, \psi_d)$. If the masses are degenerate, the theory
has a global $U(2)$ flavor symmetry.  

A finite isospin chemical potential $\mu$ is introduced as the source of the operator
\begin{equation}
  \label{eq:7}
  J^3_0\propto \bar\psi\,\tau^3\gamma_0\,\psi+\phi\,\tau^3\del_0\,\phi=n_u-n_d\,,  
\end{equation}
where $n_{u/d}$ is the charge density of the isospin fields,
$(\phi_u,\phi_d)=\phi$ and $(\psi_u,\psi_d)=\psi$. $\tau^i$ are the usual Pauli matrices. A non-zero vev
$\langle J^3_0\rangle$ introduces an isospin density as discussed in
\cite{Erdmenger:2008yj}. The isospin chemical potential $\mu$ explicitly breaks the $U(2)\simeq U(1)_B\times SU(2)_I$ flavor symmetry down to $U(1)_B\times U(1)_3$, where
$U(1)_3$ is generated by the unbroken generator $\tau^3$ of the $SU(2)_I$.
Under the $U(1)_3$ symmetry the fields with index $u$ and $d$ have 
positive and negative charge, respectively. 

In this letter we show that above a critical value for the isospin density
the system is stabilized by a $\vrho$-meson superfluid, i.e. a state with a
non-vanishing vev of the flavor current component
\begin{equation}
  \label{eq:1}
  \begin{split}
    J^1_3&\propto \bar\psi\,\tau^1\gamma_3\,\psi+\phi\,\tau^1\del_3\,\phi\\
    &=\bar\psi_u\,\gamma_3\, \psi_d+\bar \psi_d\,\gamma_3\, \psi_u+\text{bosons} \, .
  \end{split}
 \end{equation}
The vev has p-wave symmetry and 
 breaks both the $SO(3)$ rotational symmetry as well as 
the remaining Abelian $U(1)_3$ flavor symmetry. The rotational $SO(3)$ is broken down 
to $SO(2)_3$, which is generated
 by rotations around the $x^3$ axis. 
These symmetries are spontaneously broken. However only the $U(1)_3$, 
which acts on the gauge field on the D7 brane probe, is a dynamical symmetry in our approach, 
since we do not consider the backreaction of the D7 brane
gauge field on the metric. This implies that we hold the metric background and thus also its symmetries fixed. 
Consequently only one Nambu-Goldstone boson is visible in our approach, namely the one 
due to spontaneous breaking of the dynamical $U(1)_3$.

Later we will compare our results to properties of superconductors:
 the condensate $\langle J^1_3\rangle$ is the analog to the Cooper pairs of the
 BCS theory. Further, the $U(1)_3$ is the analog to the $U(1)_{\text{em}}$
and the current $J^3$ to the electric current $J_{\text{em}}$.

\section{HOLOGRAPHIC SETUP}
We consider asymptotically $AdS_5\times S^5$ spacetime which is
holographically dual to $\caln=4$ super-Yang-Mills theory with gauge group
$SU(N_c)$. The dual description of a finite temperature theory is an AdS black
hole \cite{Witten:1998zw}. We use the coordinates of \cite{Kobayashi:2006sb},
\begin{equation}
  \label{eq:bgmetric}
  \dd s^2=\frac{\vrho^2}{2R^2}
  \left(-\frac{f^2}{\ft}\dd t^2+\ft\dd\vec{x}^2\right)+\frac{R^2}{\vrho^2}
  \left(\dd\vrho^2+\vrho^2\dd\Omega_5^2\right)\,,
\end{equation}
with
\begin{equation}
  \label{eq:fft}
  f=1-\frac{\vrho_H^4}{\vrho^4}\,,\quad\ft=1+\frac{\vrho_H^4}{\vrho^4}\,,
\end{equation}
where $R$ is the AdS radius and $\dd\Omega_5$ the metric of the unit 5-sphere.
 
Flavor degrees are added by embedding $N_f$ coincident probe D7 branes
with $N_f \ll N_c$ into the AdS black hole background \cite{Karch:2002sh,
  Kruczenski:2003be, Babington:2003vm}. Here we restrict ourselves to the
case of zero quark mass and   
only comment on our massive results. In the massive case the physical
interpretation does not change. For zero quark mass the induced metric $G$ on the
D$7$-branes is 
\begin{equation}
  \label{eq:inducedmetric}
  \dd s^2(G)=\frac{\vrho^2}{2R^2}
  \left(-\frac{f^2}{\ft}\dd t^2+\ft\dd\vec{x}^2\right)+\frac{R^2}{\vrho^2}\dd\vrho^2+R^2\dd\Omega_3^2\,.
\end{equation}
The Dirac-Born-Infeld (DBI) action determines the embedding of D7-branes as
well as the gauge field on these branes,
\begin{equation}\label{eq:dbi}    
  S_{\text{DBI}}=-T_{D7}\int\!\dd^8\xi\:\str\sqrt{\left|\det\left(G+2\pi\alpha'F\right)\right|}\,,
\end{equation}
with non-Abelian gauge field $F=\dd A+c\lambda^{-1/2}\:[A,A]$ on the D$7$-branes. 
The symmetrized trace prescription in this DBI action is 
only valid to fourth order in $\alpha'$. However the corrections to the higher
order terms are suppressed by $N_f^{-1}$. Here we use two different approaches
to evaluate \eqref{eq:dbi}. First, we expand the DBI action to
fourth order. Second, to make the calculation of the full action
\eqref{eq:dbi} feasible, we modify the symmetrized trace prescription by
omitting the commutators of the generators $\tau^i$ and setting
$(\tau^i)^2=1$. The numerics presented are obtained by the second prescription.
However for both approaches we find properties of a superconductor, the
critical exponent of one half up to an error of ten percent and a second
order phase transition. The numerical values
of thermodynamical quantities differ in the two approaches due to the
modification of the effective action for the meson interaction. Still, the
qualitative behavior remains the same.

In the following we restrict to $N_f=2$ flavors. An isospin chemical
potential $\mu$ may be introduced by a non-vanishing time component of the
non-Abelian gauge field $A_0 = A_0^3 (\varrho) \, \tau^3$  on the D7-brane
probe \cite{Erdmenger:2008yj}. The operator $J_3^1$ (see \eqref{eq:1}) is
dual to the non-vanishing gauge field component $A_3^1(\varrho)$ on the D7 branes.  

The equations of motion for $A^3_0 (\varrho)$ and $A^1_3 (\varrho)$ obtained from the DBI
action \eqref{eq:dbi} are only satisfied if the asymptotic expansion near the
boundary is of the form
\begin{equation}
  \label{eq:Asym}
  \begin{split}
    A_0^3=\mu-\frac{\dt^3_0}{2\pi\alpha'}\frac{\vrho_H}{\rho^2}+\cdots\,,\;
    A_3^1=-\frac{\dt^1_3}{2\pi\alpha'}\frac{\vrho_H}{\rho^2}+\cdots\,,
  \end{split}
\end{equation}
where $\rho=\vrho/\vrho_H$ is the dimensionless AdS radial coordinate.
$\vrho_H$ is the radius of the horizon.
According to the standard AdS/CFT correspondence, 
the absence of a constant term in the expression for
  $A_3^1$ above ensures that the corresponding symmetry $U(1)_3$ is broken
  spontaneously and not explicitly. Moreover,
according to AdS/CFT, $\mu$ is the isospin chemical potential and the parameters $\dt$ are related to the vev of
the currents $J$ by
\begin{equation}
  \label{eq:density}
  \dt^3_0=\frac{2^{\frac{5}{2}}\langle J^3_0\rangle}{N_fN_c\sqrt{\lambda}T^3}\,,\quad
  \dt^1_3=\frac{2^{\frac{5}{2}}\langle J^1_3\rangle}{N_fN_c\sqrt{\lambda}T^3}\,.
\end{equation}

\subsection{Legendre transformation}

We Legendre-transform the action which gives rise to first order equations of
motion which are easier to handle numerically. The conjugate momenta are defined by
\begin{equation}
  \label{eq:2}
  p^3_0=\frac{\delta S_{\text{DBI}}}{\delta
  \left(\del_\vrho A^3_0\right)}\,,\qquad   p^1_3=\frac{\delta S_{\text{DBI}}}{\delta
  \left(\del_\vrho A^1_3\right)}\,.
\end{equation}
These conjugate momenta are functions of the AdS radial coordinate $\vrho$
--- and not constant in contrast to
\cite{Erdmenger:2008yj,Kobayashi:2006sb,Mateos:2007vc} --- due to the non-Abelian
term $A^1_3A^3_0$ in the DBI action.

Using the asymptotics \eqref{eq:Asym}, the densities $\dt$ may be
calculated from the asymptotic value of the dimensionless conjugate momenta
$\pt=p/(2\pi\alpha'N_fT_{D7}\vrho_H^3)$,   $\pt^3_0\to\dt^3_0$ and $\pt^1_3\to-\dt^1_3$.
The Legendre-transformed action is given by
\begin{equation}
  \label{eq:8}
  \tilde S_{\text{DBI}}=-N_fT_{D7}\int\dd^8\xi\sqrt{-G}
  \Bigg[
    \left(1-\frac{2c^2(\At^3_0\At^1_3)^2}{\pi^2\rho^4 f^2}\right)\left(1+\frac{8(\pt^3_0)^2}{\rho^6\ft^3}-\frac{8(\pt^1_3)^2}{\rho^6\ft
        f^2} \right)\Bigg]^{\frac{1}{2}}\,.
\end{equation}
The action simplifies to the result of~\cite{Erdmenger:2008yj} if $A^1_3\equiv 0$.
A non-zero gauge field $A^1_3$, which also implies a non-zero conjugate
momentum $p^1_3$, decreases the value of the Legendre-transformed action $\tilde
S_{\text{DBI}}$. Since the contribution of the flavors to the free energy
$F_7$ is given by 
\begin{equation}
  \label{eq:freeenergy}
  F_7=T\tilde S_{\text{DBI}}\,,
\end{equation}
the phase with non-zero gauge field $A^1_3$ is thermodynamically
favored. Since the action $\tilde S_{\text{DBI}}$ diverges on-shell due to the
infinite AdS volume, the action must be renormalized. Details may be found
in~\cite{Karch:2005ms,Erdmenger:2008yj}.

\section{PHASE TRANSITION}
\begin{figure}
  \centering
  \includegraphics[width=0.5\linewidth]{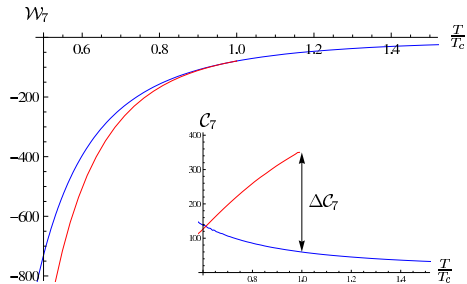}
  \caption{Dimensionless grand potential $\calw_7$ (large plot) and specific heat
    $\calc_7$ (small plot) versus temperature $T/T_c$. Blue: the unstable
    phase with $A^1_3\equiv 0$. Red: the new stable phase with $A^1_3\not\equiv 0$.
    Note that the dimensionful specific heat $C_7 \propto T^3 \calc_7$ is rising above $T_c$.}
  \label{fig:thermo}
\end{figure}

In this section we study the second order phase transition to the phase with
non-zero gauge field $A^1_3$. We work in the grand-canonical ensemble. The
contribution of the flavors to the grand potential is given by the on-shell
action $S_{\text{DBI}}$ 
\begin{equation}
  \label{eq:grandpot}
  \Omega_7=\frac{\lambda N_cN_fV_3T^4}{32}\calw_7=T S_{\text{{DBI}}}\,,
\end{equation}
where the dimensionless quantity $\calw_7$ is plotted in fig.~\ref{fig:thermo}. In
this figure we also present the contribution of the flavors to the specific heat $C_7$,
\begin{equation}
  \label{eq:specificheat}
  C_7=\frac{\lambda N_cN_fV_3T^3}{32}\calc_7=T\frac{\del^2\Omega_7}{\del T^2}\,.
\end{equation}
The temperature scale is defined by
\begin{equation}
  \label{eq:5}
  \frac{T}{T_c}=\frac{\mut_c}{\mut}=
  \left(\frac{(\dt^3_0)_c}{\dt^3_0}\right)^{\frac{1}{3}}\,,
\end{equation}
with the dimensionless critical chemical potential $\mut_c\approx 2.85$ and
density $(\dt^3_0)_c\approx 20.7$. Fig.~\ref{fig:thermo} shows a smooth
transition in the grand potential and a discontinuous step in the specific heat
at the phase transition. This implies that the transition is second order. 

From a fit to the numerical result for
the order parameter $\dt^1_3$ near the critical density, we
obtain the critical exponent of the transition to be $1/2$ up to an error
  of 10\%, in agreement with the mean field theory result. Since the order parameter 
$\dt^1_3$ and the density $\dt^3_0$ increase rapidly as $T\to 0$, we expect
that the probe approximation ignoring the backreaction of the D branes to the
geometry breaks down near $T=0$.

Let us compare our results to QCD. In QCD, the pion condensate is of
course the natural state in isospin asymmetric matter. The condensation of a
particle sets in if the isospin 
chemical potential is larger than the mass of this particle. According
to this rule, the pions condense first in QCD since they are the
Nambu-Goldstone bosons of the spontaneous chiral symmetry breaking and 
therefore the lightest particles. However the dual field theory which we
consider in this letter is supersymmetric at zero temperature and
therefore chiral symmetry cannot be broken spontaneously. In this
supersymmetric theory, the vector and scalar mesons have the same mass
at zero temperature. Due to finite temperature effects, the mass of the
vector and scalar mesons can become different as we increase the
temperature. It is a priori unclear which particle will condense. In our model we checked that the vector mesons condense
first such that the $\vrho$-meson condensation state, which we consider in
this letter, is the physical ground state of our system near the phase transition.

Since the system shares similarities with superconductors, it is
interesting to investigate the remnant of the Meissner-Ochsenfeld effect.
Switching on a magnetic field $H_3^3=F_{12}^3$ along the lines 
of~\cite{Filev:2007gb,Erdmenger:2007bn}, we find a critical   
line in the magnetic field and temperature plane. This line separates the normal phase
from the superconducting one at finite external magnetic field. Similarly to the discussion of~\cite{Hartnoll:2008kx}, this suggests the existence of a Meissner phase at small magnetic field 
and temperature.

\section{STRING THEORY PICTURE}

In the string context, the non-zero fields $A^3_0$ and $A^1_3$ induce two non-zero
flavorelectric $SU(2)$ fields $E^2_3=F^2_{03}=A^3_0 A^1_3$ and
$E^3_\vrho=F^3_{0\vrho}=-\del_\vrho A^3_0$ as well as a non-zero flavormagnetic field
$B^1_{3\vrho}=F^1_{3\vrho}=-\del_\vrho A^1_3$ on the D$7$-branes (we use the notation
of \cite{Kobayashi:2006sb,Erdmenger:2008yj} with $\varrho$ being
related to the radial AdS direction). These fields are generated
by D$7$-D$7$ strings stretched between the two probe branes and by 
strings stretched from the D$7$-branes to the horizon. The 
D$7$-D$7$ strings are new to our setup. They move from
the horizon into the bulk and thus distribute the isospin charge. This
stabilizes the system.
\begin{figure}
  \centering
  \includegraphics[width=0.5\linewidth]{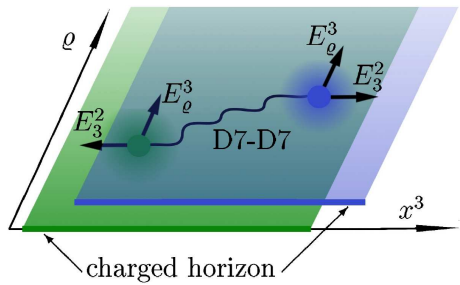}
  \caption{Sketch of our string setup: 
Strings spanned from the horizon of the AdS
    black hole to the D$7$-branes (green and blue plane) induce 
    a charge at the horizon
    \cite{Erdmenger:2008yj,Kobayashi:2006sb,Karch:2007br}. In the
    setup considered here, 
    there are also D$7$-D$7$ strings present as shown in the figure. 
    These D$7$-D$7$ strings are distributed along the AdS radial
    coordinate $\vrho$, since they have to balance the flavorelectric and
    gravitational, i.e. tension forces.  Thus these D$7$-D$7$ strings distribute
    the charges along the AdS radial coordinate, leading to a
    stable configuration of reduced energy. This corresponds to a
    superconducting condensate given by the Cooper pairs.}
  \label{fig:stringpic}
\end{figure}

Let us first describe the unstable configuration in absence of the field $A^1_3$.
As known from~\cite{Karch:2007br,Kobayashi:2006sb,Erdmenger:2008yj}, the
non-zero field $A^3_0$ is induced by fundamental strings which are stretched
from the D$7$-brane to the horizon of the black hole. Since the 
tension of these strings  would increase as they move to the boundary, they are
localized at the horizon, \ie the horizon is effectively charged. By increasing their
density, the charge on the D7-brane at the horizon and therefore the energy
of the system grows. In~\cite{Erdmenger:2008yj}, the critical 
density was found beyond which this setup becomes unstable. In this case, the
strings would prefer to move towards the boundary due to the repulsive
force on their charged endpoints generated by the field $E^3_\vrho$. 

The setup is now stabilized by the new non-zero field $A^1_3$.
This field is induced by D$7$-D$7$ strings moving in the $x^3$ direction. This
may be interpreted as a current in $x^3$ direction which induces  the magnetic
field $B^1_{3\vrho}=-\del_\vrho A^1_3$. 

Let us now explain the mechanism by which the D$7$-D$7$ strings may propagate
into the bulk. Due to the non-Abelian structure, the field $E^3_\vrho$ and the
magnetic field $B^1_{3\vrho}$ induce the field $E^2_3$, corresponding to an interaction 
between the two string types. This field $E^2_3$ 
stretches the D$7$-D$7$ strings in the $x^3$ direction. The position 
of the string in the $\vrho$
coordinate is fixed such that the gravitational force induced by the change in
tension balances the force induced by the field $E^3_\vrho$. This means that
the energy of the setup is minimized. Our numerical calculations show that
this is the case.

The double importance of the D$7$-D$7$ strings is given by the fact that they
are both responsible for stabilizing the new phase by lowering the charge
density, as well as being the dual of the Cooper pairs since they break the
$U(1)_3$ symmetry.

\section{FLUCTUATIONS}
The full gauge field~$\hat A$ on the branes consists of the field~$A$ 
and fluctuations~$a$, 
\begin{equation} 
\hat A = A_0^3 \tau^3 \dd t +A_3^1 \tau^1 \dd x_3 +
  a_\mu^a \tau^a \dd x^\mu \,,
\end{equation}
where $\tau^a$ are the $SU(2)$ generators. The linearized equations of motion
for the fluctuations $a$ are obtained by expanding the DBI action in $a$ to
second order. We will analyze the fluctuations $a^3_2$ and
$X=a^1_2+\ii a^2_2$, $Y=a^1_2-\ii a^2_2$.  

\subsection{Fluctuations in $a^3_2$}

We calculate the frequency-dependent conductivity $\sigma(\omega)$ using the Kubo formula,
\begin{equation}
  \label{eq:4}
  \sigma(\omega)=\frac{\ii}{\omega}G^R(\omega,q=0)\,,
\end{equation}
where $G^R$ is the retarded Green function of the current $J^3_2$ dual to
the fluctuation $a^3_2$, which we calculate using the method obtained
in~\cite{Son:2002sd}. The current $J^3_2$ is the analog to the electric
current since it is charged under the $U(1)_3$ symmetry. In real space it is
transverse to the condensate. Since this fluctuation is the only one which
transforms as a vector under the $SO(2)$ rotational symmetry, it decouples
from the other fluctuations of the system.
\begin{figure}
  \centering
  \includegraphics[width=0.5\linewidth]{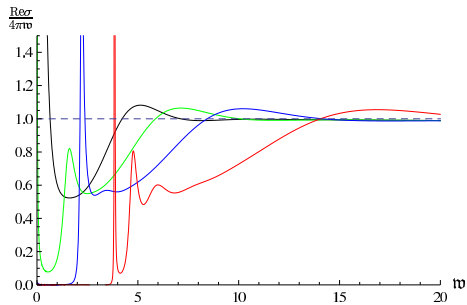}
  \caption{Real part of the conductivity $\re\sigma$ versus frequency
    $\w=\omega/(2\pi T)$ at different temperatures: $T\approx 0.90T_c$ (black), $T\approx
    0.66T_c$ (green), $T\approx 0.46T_c$  (blue), $T\approx 0.28 T_c$ (red).}
  \label{fig:resigma}
\end{figure}
The real part of the frequency-dependent conductivity
$\re\sigma(\omega)$ is presented in fig.~\ref{fig:resigma}. It shows the appearance
and growth of a gap as we increase the condensate $\dt^1_3$. Using the
Kramers-Kronig relation, which connects the real and imaginary part of the
complex conductivity, we find a delta peak at $\omega=0$ in the real part of
the conductivity, $  \re\sigma(\omega)\sim\pi n_s\delta(\omega)$.
As expected from Ginzburg-Landau theory, our numerics show that the superconductive density
$n_s$ vanishes linearly at the critical temperature, $n_s\propto (1-T/T_c)$ for
$T\approx T_c$. As a second distinct effect fig.~\ref{fig:resigma} shows 
prominent peaks which can be interpreted as mesonic excitations,   
as confirmed by our massive calculation.
This is reminiscent of results for condensed matter systems where prominent quasiparticle peaks appear
(\eg \cite{RevModPhys.46.587}).

\subsection{Fluctuations in $X=a^1_2+\ii a^2_2$, $Y=a^1_2-\ii a^2_2$}

\begin{figure}
  \centering
  \includegraphics[width=0.5\linewidth]{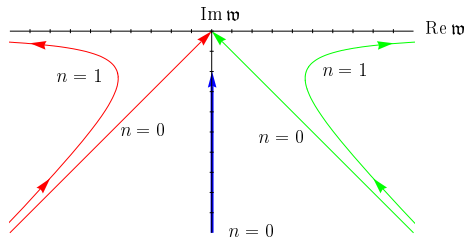}
  \caption{Movement of quasinormal modes at increasing temperature $T$:
    The different colors indicate the different fluctuations ${\color{red}X}$,
    ${\color{green}Y}$ and ${\color{blue}a^3_2}$. A Higgs mechanism is evident as explained in the text.}
  \label{fig:qnm}
\end{figure}

As shown in fig.~\ref{fig:qnm}, our setup is stable with respect to the
fluctuations $X$ and $Y$. Furthermore, fig.~\ref{fig:qnm} shows that the
quasinormal modes of higher excitations $n>1$ move to larger frequencies and
closer to the real axis. This corresponds to the formation of 
stable massive mesons. Such a behavior is known in gauge/gravity duality for
mesons which are built from massive quarks (\eg\cite{Erdmenger:2007ja}).
Thus we observe a dynamical mass generation for the mesons which we expect to
be similar to the Higgs mechanism in the field theory. This is dual to the
gravity gauge fields eating the Nambu-Goldstone bosons. This reasoning also
applies to the massive mesonic excitations in the fluctuations~$a^3_2$. The
explicit identification of the Nambu-Goldstone bosons and the massive vector mesons
is postponed to future investigations. 

\section{DISCUSSION AND OUTLOOK}
We found a stringy realization of holographic
superconductivity for which the field theory action is known. 
It will be interesting to study the D$7$-D$7$ string condensate further, 
\eg the drag force of strings pulled through this
condensate. We expect that in the condensate this force vanishes in contrast to
the original black hole background \cite{Herzog:2006gh}. 
The methods presented in this letter can also 
be applied to D$2$/D$6$ or D$3$/D$5$ (see also \cite{Roberts:2008ns}) systems.

\paragraph{ACKNOWLEDGEMENTS}
\addcontentsline{toc}{section}{Acknowledgments}
We are grateful to T. Dahm, S.~Gubser, C.~Herzog, R.~Meyer, 
A. O'Bannon, S.~Pufu and 
F.~Rust for discussions. This work was supported in part by 
{\it The Cluster of Excellence for
  Fundamental Physics - Origin and Structure of the Universe}.


\bibliographystyle{elsarticle-num}


\end{document}